\documentclass[twocolumn,showpacs,preprintnumbers,amsmath,amssymb]{revtex4}

\usepackage{graphicx}
\usepackage{dcolumn}
\usepackage{bm}

\begin{document}

\title{Particle-hole symmetric Luttinger liquids in a quantum Hall circuit}
\author{Stefano Roddaro}
\author{Vittorio Pellegrini}
\author{Fabio Beltram}
\affiliation{NEST-INFM, Scuola Normale Superiore, Piazza dei Cavalieri 7, I-56126 Pisa, Italy}
\author{Loren N. Pfeiffer}
\author{Ken W. West}
\affiliation{Bell Laboratories Lucent Technologies, Murray Hill, NJ (USA)}

\date{\today}

\begin{abstract}
We report current transmission data through a split-gate constriction
fabricated onto a two-dimensional electron system in the integer quantum
Hall (QH) regime. Split-gate biasing drives inter-edge
backscattering and is shown to lead to suppressed or enhanced
transmission, in marked contrast with the expected linear Fermi-liquid
behavior.
This evolution is described in terms of particle-hole symmetry and allows us
to conclude that an unexpected class of gate-controlled particle-hole-symmetric
chiral Luttinger Liquids (CLLs) can exist at the edges of our QH circuit.
These results highlight the role of particle-hole symmetry on the properties
of CLL edge states.
\end{abstract}

\maketitle
Quantum Hall (QH) states~\cite{Haldane} are created at integer and peculiar fractional values of the filling factor $\nu$, defined as the ratio between the electron density $n$ and the magnetic flux density $n_\phi$ measured in units of $\phi_0 = h/e$. Charge excitations confined at the edge are the only charged modes that can propagate in the QH phase along the direction set by the external magnetic field. These edge excitations at the fractional filling factor $\nu=1/m$, with $m$ odd integer, form a one-dimensional liquid that was predicted to be equivalent to a CLL~\cite{BookQH} with interaction parameter $g=\nu$~\cite{Wen,KaneFisher,Fendley}. These predictions were tested by a large number of experiments~\cite{Picciotto,Milliken,Chang,Grayson,Saminadayar,Maasilta,Roddaro} even if many open issues remain, in particular for the case of the edge states at $\nu\neq1/m$. 

\begin{figure}[h!]
\includegraphics[width=0.35\textwidth]{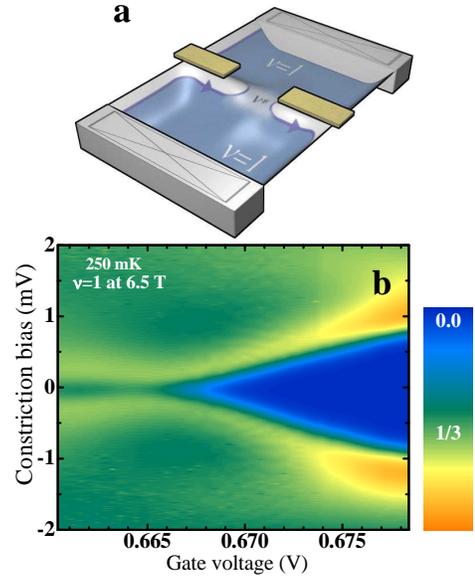}
\caption{\label{fig:1} (a) The constriction is obtained by a metallic split-gate deposited on the surface of the semiconductor. 
After application of a perpendicular magnetic field, a QH state with filling factor $\nu=1$ is formed. 
Chiral edge states that propagate along the borders of the sample are brought in close proximity by the 
constriction when the metal gates are negatively biased. In this way a local backscattering between edge states is induced. 
In addition, local depletion of the 2DES close to the metal gates creates a region with a reduced filling factor $\nu^*$. 
(b) Finite bias transmission curves ($t=hG/e^2$) as a function of split-gate voltage at $T=250\,{\rm mK}$ and $\nu=1$. 
The constriction bias $V$ is defined as the voltage difference between the two interacting edge states. }
\end{figure}

A split-gate (SG) technique~\cite{Thornton} can be exploited to define a nanofabricated constriction in order to induce a 
controllable scattering between counter-propagating edge channels that are locally brought in close proximity (see Fig.1a). 
The constriction thus realizes an artificial impurity and can be used to test one of the most significant manifestations of 
CLL behavior: the complete suppression of the (low-temperature, low-bias) transmission through the impurity and its related 
power-law behavior~\cite{Wen,KaneFisher,Fendley}. Backscattering at the constriction is controlled by the 
split-gate voltage $V_g$: by increasing $|V_g|$ the inter-edge distance is decreased; at larger $|V_g|$ values, in addition, 
the density of the two-dimensional electron system in proximity to the SG is appreciably reduced. In the presence of a 
uniform external magnetic field, this leads to a reduced filling factor $\nu^*$ within the constriction region (see Fig.~\ref{fig:1}a). 

Here we show that the SG voltage $V_g$ not only modifies the backscattering strength but also defines unexpected 
robust CLLs that are related by particle-hole symmetry. In order to demonstrate this we study the 
constriction transmission in the QH regime at bulk {\em integer} filling factor $\nu=1$. 
The measured low-energy conductance displays a non-linear behavior determined by the SG voltage. 
Both suppression and enhancement of the transmission through the constriction were observed, in disagreement 
with the usual expectation of linear tunneling for $g=\nu=1$. We argue that these data reveal the 
existence of peculiar CLLs with an effective $g$ determined by the SG voltage 
and that the intrinsic particle-hole symmetry~\cite{Girvin,MacDonald} of the QH system
influences the link between edge states and CLLs.

Devices were realized starting from a high-mobility ($4.6\times10^6\,{\rm cm^2/Vs}$) AlGaAs/GaAs single heterojunciton. The two-dimensional electron system (2DES) was buried $100\,{\rm nm}$ below the surface, with a sheet density of $1.5\times10^{11}\,{\rm cm^{-2}}$. Gates were realized by thermal evaporation of a metal bilayer Al/Au ($5/25\,{\rm nm}$) with different constriction gaps ranging between $600$ and $800\,{\rm nm}$. Measurements were performed in a $^3$He refrigerator with base temperature of $250\,{\rm mK}$. The filling factor was kept to $\nu=1$ by tuning the magnetic field (at about $6.5\,{\rm T}$) to the center of the vanishing longitudinal resistivity $\rho_{xx}$ measured far form the constriction. We measured the transmission amplitude $t=hG/e^2$ ($G$ is the differential conductance) as a function of the negative SG voltage $V_g$. The transmission coefficient $t$ was obtained from finite bias differential conductance measurements based on a standard two-wire phase-locked technique.

\begin{figure}[h!]
\includegraphics[width=0.4\textwidth]{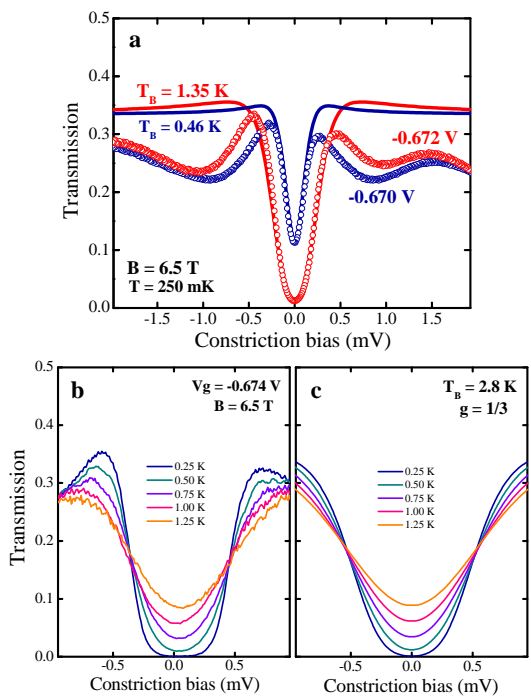}
\caption{\label{fig:2} (a) Comparison between experimental (circles) and theoretical (lines) transmission characteristics calculated from the CLL theory of inter-edge tunneling at $g=1/3$ as developed in ref.~\cite{Fendley}. The width of the dip before saturation is in {\em parameter-free} agreement with theory. The saturation to $t=0$ is well-described by adjusting the single fit parameter of the tunneling strength $T_B$. (b) Evolution of finite bias transmission curve at $V_g=-0.674\,{\rm V}$ as a function of temperature. (c) CLL theory prediction of the temperature dependence of the transmission at $g=1/3$. Agreement is obtained with a {\em single} adjustable fit parameter ($T_B=2.8\,{\rm K}$) for {\em all} curves.}
\end{figure}

Figure~\ref{fig:1}b reports a color-plot of the measured transmission characteristics at different values of $V_g$ close to SG pinch-off. The curves display a marked non-linearity and evolve from a small zero-bias suppression of the transmission ($V_g\approx-0.665\,{\rm V}$) to a deep minimum that saturates at zero (blue central region). When $g=\nu=1$, the edge channel should behave like a Fermi liquid and linear tunneling is expected. The observed non-linear behavior indicates that the relevant interaction $g$ is not equal to $1$ but is determined by the SG voltage, confirming similar results obtained in the fractional QH regime~\cite{Roddaro,Chung}. We now show that the effective interaction is determined by the SG voltage through the creation of a region in which a fractional QH state emerges. Figure~\ref{fig:2}a reports a representative set of experimental curves in which at higher bias $t\approx1/3$. This suggests that in the constriction region $\nu<1$ and inter-edge tunneling is dominated by scattering between edge states in the reduced filling $\nu^*=1/3$. Consistently with this interpretation the experimental data are successfully described by the theory introduced by Fendely {\em et al.}~\cite{Fendley} for CLLs with $g=1/3$. Notably, when the transmission dip is small (i.e. far from the pinch-off limit $t=0$), the non-linearity width is in parameter-free agreement with this theory. The $V_g$ evolution can be captured by using a single adjustable parameter: the so-called impurity (or constriction) interaction strength $T_B$. CLL $g=1/3$ behavior is also confirmed by the temperature evolution. Figures~\ref{fig:2}b,c report experimental and calculated transmission curves at different temperatures. In the calculations the interaction strength was set to $T_B=2.8\,{\rm K}$ in order to fit the measured $V=0$ transmission values. This one adjustment yields agreement in the whole voltage-bias and temperature range tested. Significantly the fixed point experimentally observed at $t\approx0.2$ is present also in the theory. 

It is remarkable and somewhat unexpected to observe a distinct CLL-behavior in a structure nominally in an integer QH configuration since the local filling factor $\nu^*$ is defined in a rather limited region of the sample while the observation of CLL characteristics should require the presence of an {\em extended} correlated fractional edge structure. One possible explanation is that the fractional-edge stripe at $\nu^*=1/3$ also extends along the smooth edge defined by the SG so that the effect of the constriction is to transmit selectively this specific edge channel. We note, in addition, that single-particle effects can be ruled out as the only available scale voltage, the Landau level gap $\hbar\omega_c \approx 10\,{\rm meV}$, is one order of magnitude bigger than the observed non-linearity range. Coulomb blockade effects are inconsistent with the observed peak width and high-bias. In order to support our interpretation of the data, it is important to point out that $\rho_{xx}$ and Hall resistivity $\rho_{xy}$ measurements performed on depleted regions of the 2DES (we adopted a $100\times100\,{\rm nm^2}$ top gate for this analysis) ensure that the observation of fractional states at $\nu^*=1/3$ and $\nu^*=2/3$ is well within sample quality, base temperature and magnetic field values. The observed non-linearity is not related to any imperfection of the measured constriction: similar data were obtained on different devices with different constriction geometries. Devices displayed a sharp quantization of the conductance at zero magnetic field, without resonant features.

\begin{widetext}

\begin{figure}[h!]
\includegraphics[width=0.6\textwidth]{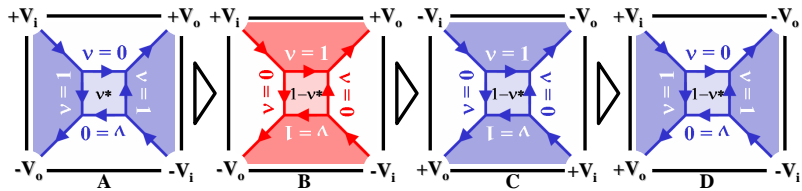}
\caption{\label{fig:3} Particle-hole symmetry in a quantum Hall circuit with constriction. Arrows indicate the direction of propagation of edge states and voltages at the corner show the potentials of ingoing and outgoing (after interaction/equilibration) edge channels. The constriction of panel (a) can be described in terms of holes (b). Thanks to particle-hole symmetry, holes in (b) follow the same dynamics of time-reversed electrons or, equivalently, of electrons in reversed external electromagnetic field (c). The circuit (c) can be mapped back to the original constriction geometry by a $180$ degrees rotation around the axis defined by outgoing edge states. This mapping establishes a correspondence between circuit (a) and (d): finite bias transmission $t(V)$ and reflection $r(V)$ swap when going from $\nu^*$ to $1-\nu^*$, i.e. $t_{\nu^*}(V)=r_{\nu^*}(V)$.}
\end{figure}

\end{widetext}

Having established that the CLL behavior is associated to the fractional QH edge states with filling factor $\nu^*$ we are now in the position to show that our circuit implements a QH geometry which is self-dual under charge-conjugation in the first Landau level. Indeed a partially filled first Landau level (with $\nu<1$) can be described either in terms of electrons or in terms of holes with filling $1-\nu$ over a completely filled Landau level. As long as inter-level mixing can be neglected, the complementary hole picture is governed by a time-reversed version of the electronic hamiltonian and a mapping between states at $\nu\leftrightarrow 1-\nu$ can be established~\cite{Girvin,MacDonald}. An extension to a space-dependent $\nu(x)\leftrightarrow 1-\nu(x)$ is possible, provided the in-plane field ${\bf E}$ does not induce inter-level mixing (second order perturbation theory requires $e|{\bf E}|\ll \hbar\omega_c/\ell_m$, where $\ell_m=\sqrt{\hbar/eB}$). To understand the impact of this particle-hole symmetry on the experiments here presented, we show in Fig.~\ref{fig:3}a a cartoon of an ideal QH constriction at ``local'' filling factor $\nu^*$ and biased at $V = 2V_i$. Chiral edge states propagate at the boundary between different filling factors in the directions indicated by the arrows: incoming edge states are polarized~\cite{Note} at $\pm V_i$ and emerge after interaction/equilibration at a potential $\pm V_o(Vi)$. The same circuit is described in terms of holes in Fig.~\ref{fig:3}b with constriction filling factor $1-\nu^*$. Following particle-hole symmetry, the hole-dynamics in Fig.~\ref{fig:3}b is the same of time-reversed electrons or, equivalently, of electrons in a reversed external electromagnetic field (Fig.~\ref{fig:3}c). A final $180$ degrees rotation (around the axis defined by outgoing edge states) maps the system back to a biased constriction with central filling $1-\nu^*$. 

This simple particle-hole argument sets a correspondence between the behavior of the QH circuit at constriction filling factor $\nu^*$ and that with $1-\nu^*$. The main consequence of the mapping is that the meaning of ``crossing the constriction'' changes in the transformation Fig.~\ref{fig:3}a $\to$ Fig.~\ref{fig:3}d: transmitted trajectories of Fig.~\ref{fig:3}a map on reflected trajectories of Fig.~\ref{fig:3}d and vice versa. This suggests that transmission $t$ amplitudes in Fig.~\ref{fig:3}a are equal to reflection r amplitudes in Fig.~\ref{fig:3}d. More specifically the currents flowing in the constrictions of Fig.~\ref{fig:3}a and Fig.~\ref{fig:3}d are

\begin{eqnarray}
I_{\nu^*}(V) &=& G_0\left(V_i+V_o\right)\\
I_{1-\nu^*}(V) &=& G_0\left(V_i-V_o\right)
\end{eqnarray}

where $G_0=e^2/h$. The sum of these two currents is $I_{1-\nu^*}(V)+I_{\nu^*}(V)=G_0V$, so that in terms of differential transmission $t=G_0^{-1}dI/dV$ and reflection $r$ we have the following duality

\begin{equation}
t_{\nu^*}(V)=1-t_{1-\nu^*}(V)=r_{1-\nu^*}(V)
\end{equation}

\begin{figure}[h!]
\includegraphics[width=0.3\textwidth]{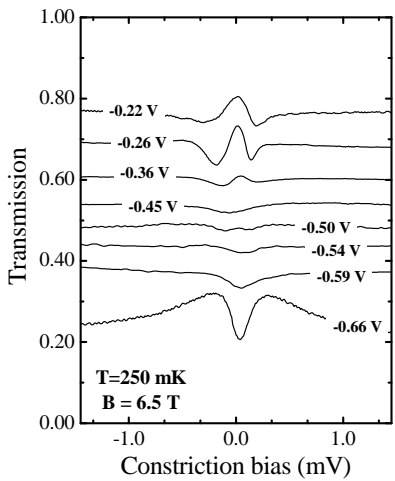}
\caption{\label{fig:4} Complete evolution of finite bias transmission as a function of the gate voltage $V_g$. Minima observed for $t\approx1/3$ evolve in a flat region at $t\approx1/2$ and finally to zero-bias peaks at $t\approx2/3$.}
\end{figure}

The circuit of Fig.~\ref{fig:3}a is therefore dual respect to the one of Fig.~\ref{fig:3}d, while the fixed point at $t=1/2$ (corresponding to $\nu^*=1/2$) realizes a self-dual configuration. Since the local filling factor $\nu^*$ can be tuned by the SG voltage, our QH circuit (Fig.~\ref{fig:1}a) allows us to test directly the occurrence of this duality in the inter-edge tunneling at the constriction and to verify the presence of unexpected CLLs. Figure~\ref{fig:4} shows the measured evolution of the transmission curves versus $V_g$ and presents a behavior that is consistent with the charge-conjugation argument presented above. Starting from the pinch-off limit, curves become flat when $|V_g|$ is reduced to a crossover point at $t\approx1/2$. A further reduction of the tunneling strength leads to symmetric curves at $t\approx2/3$ characterized by an {\em enhanced} zero-bias transmission. This behavior was verified on a set of different constrictions and crossover of the transmission was always found consistent with $t=1/2$. Curves with a zero-bias transmission enhancement seem to contradict any available CLL prediction for $g<1$. This behavior, however, can be explained following the charge-conjugation argument. Let us consider for instance the $\nu^*=2/3$ case. The electronic circuit can be analyzed in terms of its dual hole system at central filling $\nu^*=1/3$ (Fig.~\ref{fig:3}a $\to$ Fig.~\ref{fig:3}b). In this dual picture, the constriction induces a backscattering between edge states at $\nu^*=1/3$, that should thus be described in terms of a CLL with $g=1/3$. The behavior of such a circuit is well-known: enhanced CLL reflection is expected at zero bias leading to increased constriction transmission. An equivalent way to explain this is to recognize that channels at the boundary of the $\nu=1$ and $\nu^*=2/3$ regions are (charge-conjugate) $g=1/3$ CLLs between which non-linear tunneling can occur. This hole-symmetric CLL tunneling is however quite peculiar: fractional quasiparticles are allowed to cross the constriction region at $\nu^*=2/3$ while only (full) electrons can be reflected. This is the exact opposite of what happens for $\nu^*=1/3$ and accounts for the observed ``reversed'' behavior in the $t>1/2$ region. Finally we would like to point out that the tunneling scheme realized at $t>1/2$ is {\em not} in contradiction with CLL transmission through the impurity. It does reflect, however, an alternative way to map the backscattering of chiral edges on the impurity problem in a non-chiral Tomonaga Luttinger liquid (TLL). While the standard approach assumes that inter-edge backscattering maps on TLL backscattering, we argue that in the $\nu^*\approx2/3$ case it rather maps on TLL forward scattering. These are two complementary and non-equivalent ways to ideally join chiral edges and form the complete non-chiral TLL. Both are realized in the $\nu=1$ circuit, where they are related by charge conjugation. Similar results were obtained at a bulk filling factor $\nu=2$ (data not shown) that display two minimum-to-maximum crossovers and linear characteristics at $t\approx1/4$, $1/2$, and $3/4$ indicative of independent effects of the two channels that constitute the $\nu=2$ edge. 

Our argument provides a simple and complete description of the experimental data and highlights the role of QH particle-hole symmetry in the correspondence between edge backscattering and TLLs. The creation and gate-voltage control of CLLs in a QH circuit like the one studied here has recently attracted significant theoretical interest. In~\cite{Papa} it was suggested that inter-edge Coulomb interaction along the split-gate (coupling electrons on the left and right sides of it) could also modify the chiral channel, leading to a CLL with either $g<1$ or $g>1$ as a function of the gate voltage $V_g$. Additional theoretical work is currently devoted to the modeling of the low-temperature suppression inter-edge backscattering~\cite{DAgosta}. Finally it would be interesting to extend the QH particle-hole symmetry to explain data taken at the fractional bulk filling factor of 1/3 \cite{Roddaro,Chung}.

We acknowledge support from the Italian Ministry of Research and from the European Community's Human Potential Program. We thank Allan H. MacDonald, Emiliano Papa, Roberto Raimondi and Giovanni Vignale for illuminating discussions.


\begin{thebibliography}{9}
\bibitem{Note} In all of the panels of Fig.3, the edge-voltages can be thought to be present at four different ohmic contacts located at the corners of the square.
\end{thebibliography}
\end{document}